\documentclass[twocolumn,twoside]{revtex4}
\usepackage{graphicx}
\usepackage{fancyhdr}
\usepackage{overpic}
\begin{document}

\title{Charmonium and Charm Spectroscopy}

%

\author{Y. Hu\\
on behalf of the BESIII collaboration}
\affiliation{Institute of High Energy Physics, Beijing 100049, China}

\begin{abstract}
In this talk, we review the recent experimental developments on charmonium and charm spectroscopy based on the data samples taken by the BESIII, Belle, LHCb and CMS experiments. We concentrate on the resonant parameter measurement of $\chi_{cJ}$, $\eta_{c}$, $\eta_{c}(2S)$, observation of $\psi(1^{3}D_{3})$ candidate and alternative $\chi_{c0}(2P)$ candidate, observation of excited $B_{c}$ states, $\Xi(2930)$ states, excited $\Omega_{c}$ states and doubly charm baryon, and the study of $\Lambda_{c}^{*}$ states.
\end{abstract}

\maketitle

\thispagestyle{fancy}


\section{Introduction}

Since the observation of $J/\psi$ meson in 1974~\cite{JPsi1974_1,JPsi1974_2}, many charmonium states have been observed~\cite{psip1974,psipp1977}. Charmonia are charmed-quark and anticharmed-quark states ($C\bar{C}$) bound by the strong interaction, and they can be described well by potential models. States below the $D\bar{D}$ threshold have all been observed, and match with prediction well. Above the $D\bar{D}$ threshold, only few states have been measured and identified. Further more, after the discovery of the $X(3872)$ meson~\cite{X3872_Belle}, a lot of unexpected states (called charmoniumlike states or XYZ particles) have been observed~\cite{XYZreview1,XYZreview2,XYZreview3}. They could be candidates for charmonium states, however, they have strange properties which make them more like exotic states rather than conventional mesons.


 Charmed baryons consist of one heavy charm quark and two light (u, d, s) quarks.  Large mass difference provides a natural way to classify these states using Heavy Quark Effective Theory (HQET). Diquark correlation is enhanced by weak Color Magnetic Interaction with a heavy quark. So diquark can be considered as new degree of freedom.   Charmed baryon spectroscopy provides an ideal place for studying the dynamics of the light quarks in the environment of a heavy quark. Recent years there are great progresses in our understandings of the charmed baryons. But the spin-parity assignments of many of the observed states are still to be discovered.

The $B_{c}$ meson is unique system of two heavy quarks in a bound state. The ground state was discovered in 1998 by the CDF Collaboration~\cite{Bc_CDF}. The spectrum of this heavy quarkonium family is predicted to be very populated by QCD potential models and Lattice QCD~\cite{Bc_pre_1,Bc_pre_2,Bc_pre_3,Bc_pre_4}, but spectroscopic observations and measurements of production properties remain scarce due to the small production rate.

In this article,  we briefly introduce the most recent results on the Charmonium and Charm Spectroscopy.

\section{Charmonium Spectroscopy}

\subsection{Precise  $\chi_{c1,2}$  parameters measurement using $\chi_{c1,2} \to J/\psi \mu^{+} \mu^{-}$}

\begin{figure}[b]
\centering
\includegraphics[width=75mm]{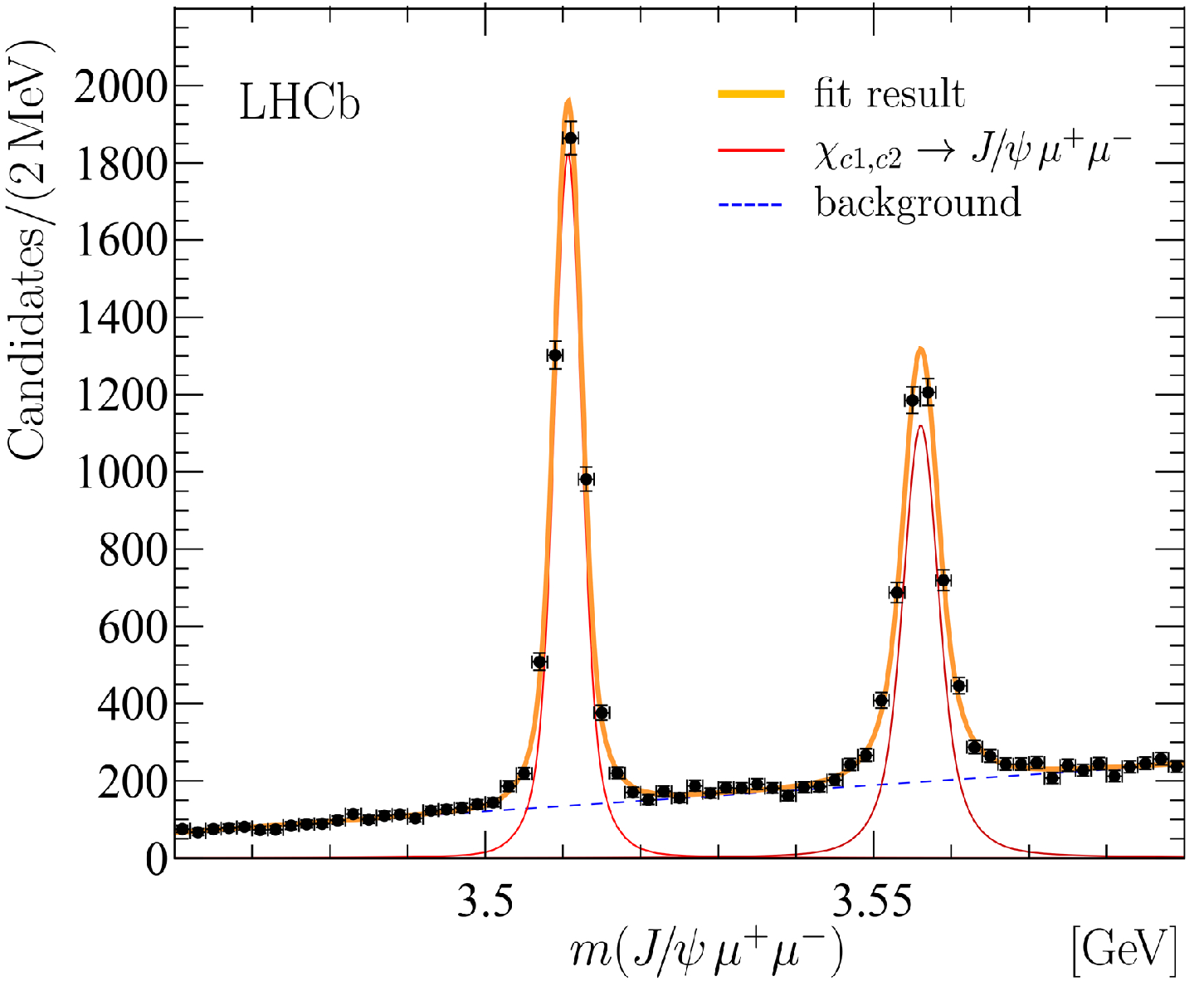}
\caption{Mass distribution for selected $J/\psi\mu^{+}\mu^{-}$ candidates.} \label{Jpsiuu}
\end{figure}

Recently, LHCb observed the decays $\chi_{c1,2} \to J/\psi \mu^{+} \mu^{-}$ base on its full run I and part of run II data~\cite{Chcj_Jpsimumu_LHCb}.  Then those decays were used to measured the $\chi_{c1}$ and $\chi_{c2}$ masses together with the $\chi_{c2}$ natural width. With an extended unbinned maximum likelihood fit  performing to the $J/\psi \mu^{+}\mu^{-}$ invariant mass distribution as shown in Fig.~\ref{Jpsiuu}, the masses of $\chi_{c1}$ and $\chi_{c2}$ are measured as 3510.71 $\pm$ 0.04 $\pm$ 0.09 MeV$/c^{2}$ and 3556.10 $\pm$ 0.06 $\pm$ 0.11 MeV$/c^{2}$, respectively, and the natural width of $\chi_{c2}$ is measured as 2.10 $\pm$ 0.20 $\pm$ 0.02 MeV. The measurements are in good agreement with and have comparable precision to the world average~\cite{PDG2018}. This observations open up a new avenue for hadron spectroscopy at the LHC. It will be possible to extend measurements down to very low $p_{T}(\chi_{c1,c2})$ probing further QCD predictions.

Soon after, BESIII also studied the decays $\chi_{c1,2} \to J/\psi \mu^{+} \mu^{-}$ via the radiative process $\psi(3686)\to\gamma\chi_{cJ}$~\cite{Chcj_Jpsimumu_BESIII}. With the world-average branching fractions of $\psi(3686)\to\gamma\chi_{cJ}$ and $J/\psi\to l^{+} l^{-}$~\cite{PDG2018}, the absolute branching fractions of $\chi_{c1,2} \to J/\psi \mu^{+} \mu^{-}$ are measured as (2.51 $\pm$ 0.18 $\pm$ 0.20) $\times 10^{-4}$ and (2.33 $\pm$ 0.18 $\pm$ 0.29) $\times 10^{-4}$, respectively. Which can be used to study the production of $\chi_{c1,2}$ with this mode on LHC experiments.


\begin{figure}[t]
\centering
\includegraphics[width=62mm]{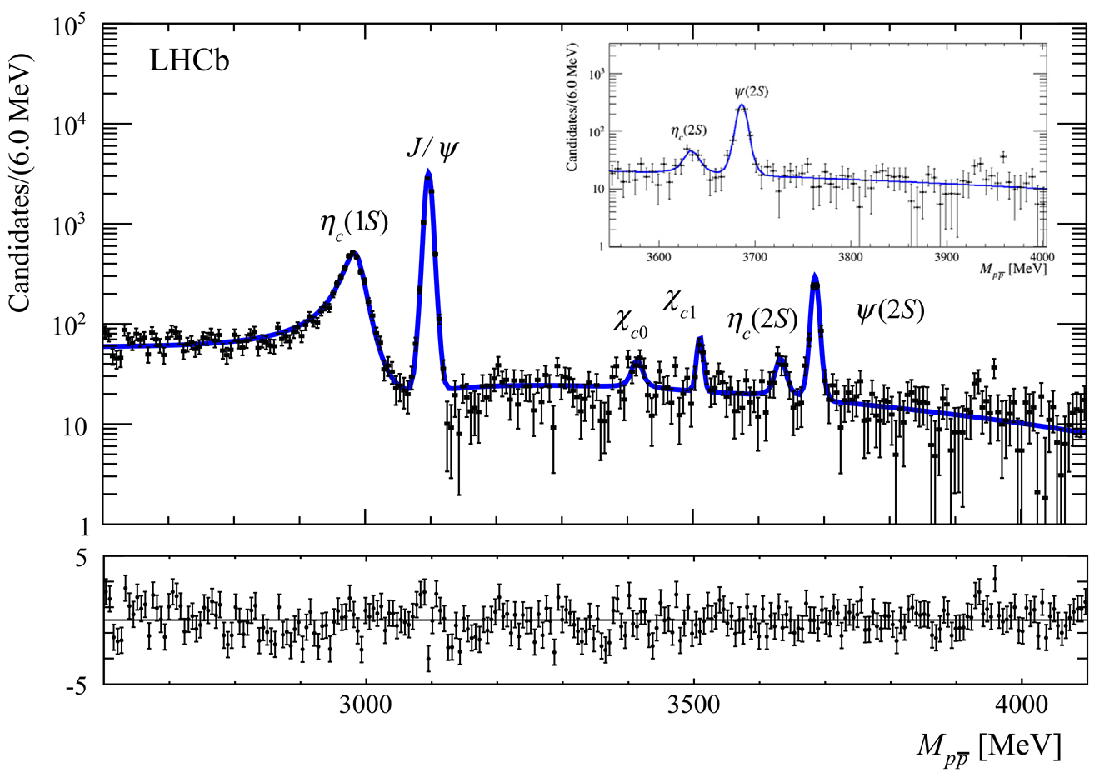}
\includegraphics[width=64mm]{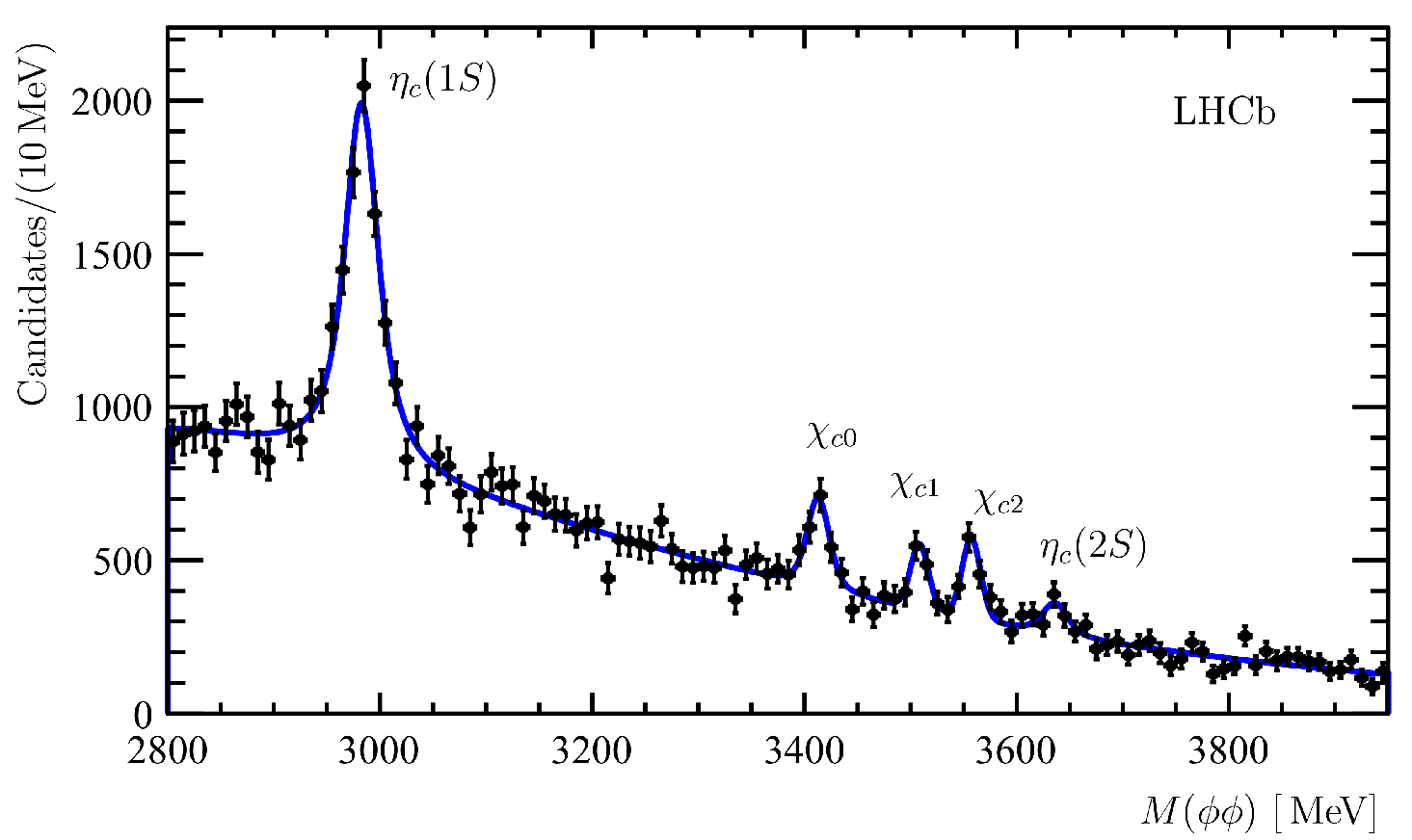}
\includegraphics[width=64mm]{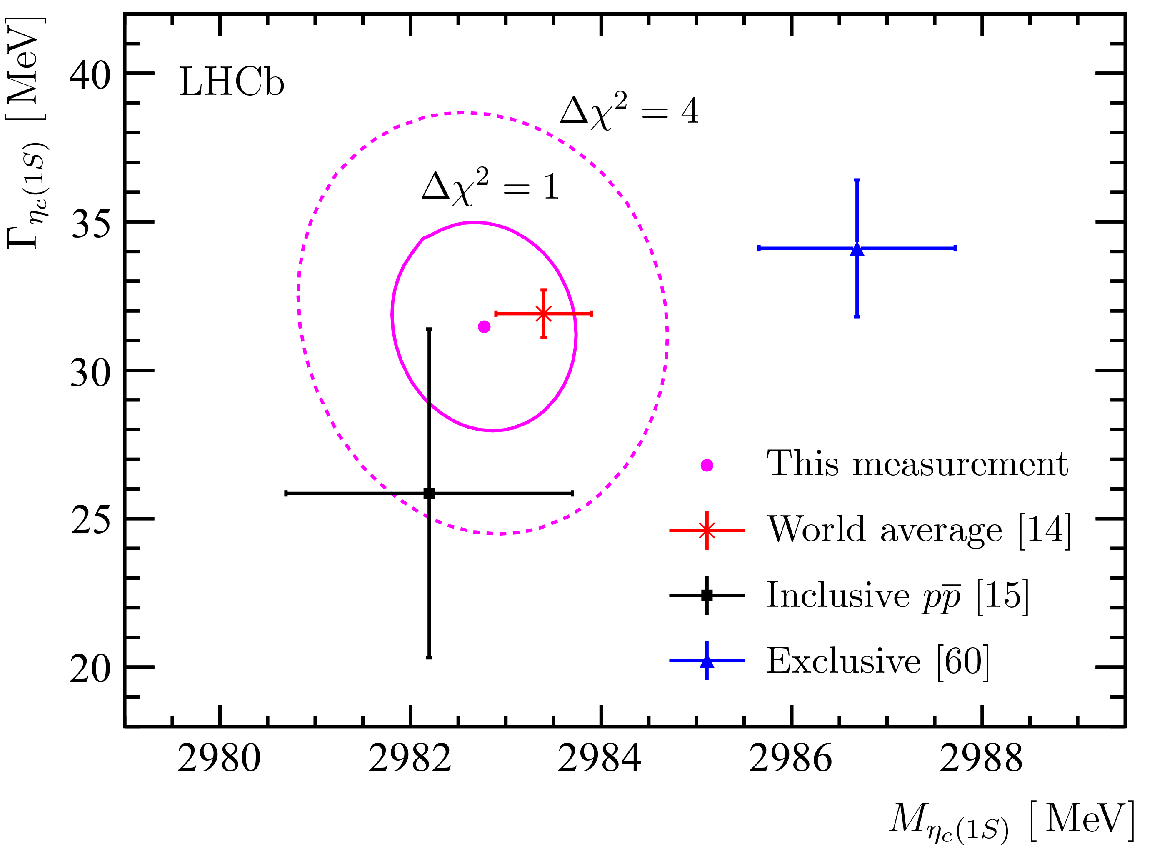}
\caption{Invariant mass spectrum of the $p\bar{p}$ candidates (top) and $\phi\phi$ combinations (middle). Contour plot of $\Gamma_{\eta_{c}(1S)}$ and $M_{\eta_{c}(1S)}$ using $\eta_{c} \to \phi\phi$ decays (bottom). The red cross, black square and blue triangle with error bars indicate the world average~\cite{PDG2018}, the result from Ref.~\cite{Etac_PP_LHCb}, and the result from Ref.~\cite{Etac_PP_LHCb_ex}, respectively.} \label{Bdecay_charmonium}
\end{figure}

\subsection{Charmonia from $B$ meson decay}

The $\eta_{c}(1S)$ state is the lowest S-wave spin-singlet charmonium state and has been observed in various processes. The measurements of the $\eta_{c}(1S)$ mass and width in radiative charmonium transitions show a tension with those determined in different processes such as two-photon production and B decays~\cite{PDG2018}.  Recently, LHCb performed a study of $B^{+}\to p\bar{p}K^{+}$ with the exclusive reconstruction method~\cite{Etac_PP_LHCb}. The $\eta_{c}(2S)$ is first observed in  $p\bar{p}$ final states as shown in Fig.~\ref{Bdecay_charmonium}. The relative branching fraction is measured as: $\mathcal{R}_{\eta_{c}(2S)} \equiv \frac{\mathcal{B}(B^{+}\rightarrow\eta_{c}(2S)K^{+})\times\mathcal{B}(\eta_{c}(2S)\rightarrow p\bar{p})}{\mathcal{B}(B^{+}\rightarrow J/\psi K^{+})\times\mathcal{B}(J/\psi\rightarrow p\bar{p})}$ = (1.58 $\pm$ 0.33 $\pm$ 0.09) $\times 10^{-2}$. But there are no signals for decays $B^{+}\to\psi(3770)(\to p\bar{p})K^{+}$ and $B^{+}\to X(3872)(\to p\bar{p})K^{+}$ observed. The upper limits of the relative branching fractions at 90 \% confidence level are estimated as: $\mathcal{R}_{\psi(3770)} < 9 \times 10^{-2}$ and $\mathcal{R}_{X(3872)} < 0.20 \times 10^{-2}$, respectively. The differences between $M(J/\psi)$ and $M(\eta_{c}(1S))$, between $M(\psi(2S))$ and $M(\eta_{c}(2S))$ are measured to be 110.2 $\pm$ 0.5 $\pm$ 0.9 MeV/$c^{2}$ and 52.5 $\pm$ 1.7 $\pm$ 0.6 MeV/$c^{2}$, respectively. The natural width of the $\eta_{c}(1S)$ is measured as $\Gamma_{\eta_{c}(1S)}$ = 34.0 $\pm$ 1.9 $\pm$ 1.3 MeV. In contrast to the determinations using radiative decays, these mass and width determinations do not depend on the knowledge of the line shapes of the magnetic dipole transition.

Soon after, LHCb performed a study of the inclusive production of charmonium states in b-hadron decays using decays to $\phi$-mesons~\cite{Etac_phiphi_LHCb}. The first evidence for the decay $\eta_{c}(2S)\to \phi\phi$ is observed as shown in Fig.~\ref{Bdecay_charmonium}. The masses of $\chi_{c0}$, $\chi_{c1}$, $\chi_{c2}$, $\eta_{c}(1S)$ and $\eta_{c}(2S)$ are measured as  3413.0 $\pm$  1.9 $\pm$  0.6 MeV$/c^{2}$, 3508.4 $\pm$  1.9 $\pm$  0.7 MeV$/c^{2}$, 3557.3 $\pm$ 1.7 $\pm$ 0.7 MeV$/c^{2}$, 2982.8 $\pm$  1.0 $\pm$  0.5 MeV$/c^{2}$ and 3636.4 $\pm$  4.1 $\pm$  0.7 MeV$/c^{2}$, respectively. The natural width of the $\eta_{c}(1S)$ is measured as 31.4 $\pm$ 3.5 $\pm$ 2.0 MeV. The measurements of the $\eta_{c}(1S)$ mass and natural width using $\eta_{c}(1S)$ meson decays to $\phi\phi$ are consistent with the studies using decays to $p\bar{p}$  and the world average. The measured $\eta_{c}(1S)$ mass is below the result in exclusive method~\cite{Etac_PP_LHCb_ex} as shown in Fig.~\ref{Bdecay_charmonium}.

\subsection{Charmonium in near-threshold $D\bar{D}$ spectroscopy}
\begin{figure}[t]
\centering
\includegraphics[width=80mm]{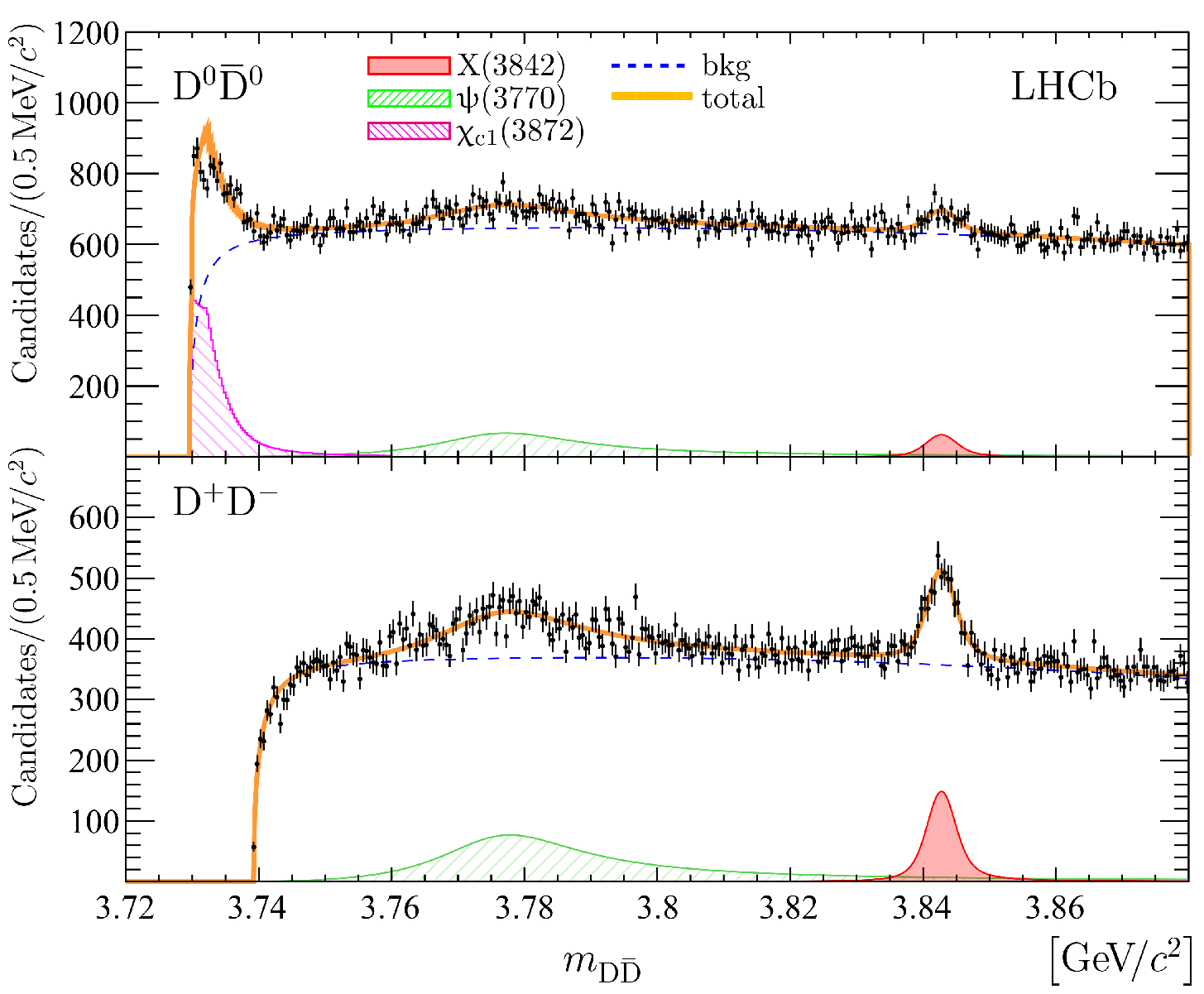}
\caption{Mass spectra of (top) $D^{0}\bar{D}^{0}$ and (bottom) $D^{+}D^{-}$ candidates in the near-threshold $m_{D\bar{D}}<$ 3.88 GeV/$c^{2}$ region.} \label{Psi3D3}
\end{figure}

 Recently, with the full Run I and Run II data, LHCb observed a new narrow charmonium state, the X(3842) resonance in $D^{0}\bar{D}^{0}$  and $D^{+}D^{-}$ final states as shown in Fig.~\ref{Psi3D3}~\cite{psi3D3_LHCb}. The mass and the natural width of this state are measured to be 3842.71 $\pm$ 0.16 $\pm$ 0.12 MeV/$c^{2}$ and 2.79 $\pm$ 0.51 $\pm$ 0.35 MeV, respectively. The observed mass and narrow natural width suggest the interpretation of the new state as the unobserved spin-3 $\psi_{3}(1^{3}D_{3})$ charmonium state. In addition, prompt hadroproduction of the $\psi(3770)$ and $\chi_{c2}(3930)$ states is observed for the first time. The mass of $\psi(3770)$ is measured as 3778.1 $\pm$ 0.7 $\pm$ 0.6 MeV/$c^{2}$, which agrees well with and has a better precision than the current world average~\cite{PDG2018}. The mass and width of $\chi_{c2}(3930)$ are measured as 3921.9 $\pm$ 0.6 $\pm$ 0.2 MeV/$c^{2}$ and 36.6 $\pm$ 1.9 $\pm$ 0.9 MeV, which are more precise than previous measurements made at $e^{+}e^{-}$ machines. But the mass is 2$\sigma$ lower than the current world average whilst the natural width is 2$\sigma$ higher. It is interesting to note that the measured value of the mass is roughly midway between the masses quoted in Ref.~\cite{TBarnes} for this state and and for the $X(3915)$ meson, which is only known to decay to the $J/\psi\omega$ final state~\cite{X3915_B_jpsiOmega_BaBar}. Further studies are needed to understand if there are two distinct charmonium states in this region or only one.

 \subsection{Alternative $\chi_{c0}(2P)$ candidate in $e^{+}e^{-}\to J/\psi D\bar{D}$}

$X(3915)$ was observed by Belle~\cite{X3915_B_jpsiOmega_Belle}, then confirmed by BaBar~\cite{X3915_B_jpsiOmega_BaBar} in $B\to(J/\psi\omega)K$. Soon after, it was observed by both Belle and BaBar in $\gamma\gamma\to J/\psi\omega$~\cite{X3915_gammagamma_jpsiOmega_Belle,X3915_gammagamma_jpsiOmega_BaBar}. BaBar determined its $J^{PC}$ as  $0^{++}$, then identified it as the candidate of $\chi_{c0}(2P)$. But this will face several difficulties: $X(3915)$ is too narrower with a measured width 20 MeV than the expected $\chi_{c0}(2P)$ width 100 MeV~\cite{Chic02P_width}; $\chi_{c0}(2P)$ is expected to decay strongly to $D\bar{D}$ in an S-wave~\cite{Barnes}, but has not yes been observed experimentally for $X(3915)$; in this case, the $2^{3}P_{2}-2^{3}P_{0}$ splitting is unnaturally smaller than the mass difference for the bottomonium states $\chi_{bJ}(2P)$, which is inconsistent with expectations based on the heavier bottom quark mass.

\begin{figure}[t]
\centering
\includegraphics[width=75mm]{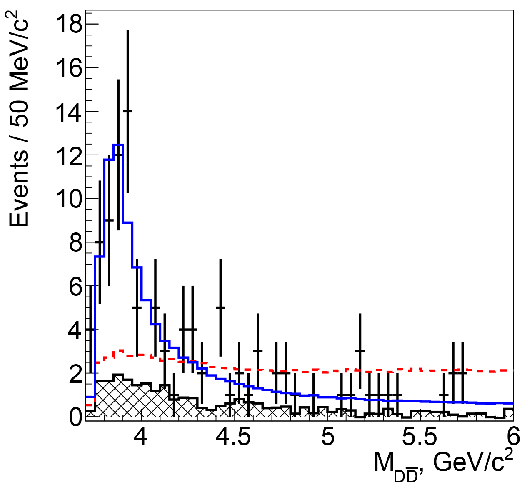}
\caption{Projections of the signal fit results in the default model onto $M_{D\bar{D}}$. The points with error bars are the data, the hatched histograms are the background, the blue solid line is the fit with a new $X^{*}$ resonance ($J^{PC} = 0^{++}$) and the red dashed line is the fit with nonresonant amplitude only.} \label{chic0_2_belle}
\end{figure}

Recently, Belle attempted to search for alternative  $\chi_{c0}(2P)$ via double-charmonium production in association with the $J/\psi$~\cite{Chic02P_Belle}. With full amplitude analysis of $e^{+}e^{-}\to J/\psi D\bar{D}$, they observed an obvious signal for a new charmoniumlike state $X^{*}(3860)$  in the $D\bar{D}$ final states as shown in Fig.~\ref{chic0_2_belle}. Its mass and width are measured as $3862 _{-32}^{+26}$$_{-13}^{+40}$ MeV/$c^{2}$ and $201_{-67}^{+154}$$_{-82}^{+88}$ MeV, respectively. Which is consistent with potential model expectations for $\chi_{c0}(2P)$~\cite{Chic02P_Theory}.  And the $J^{PC} = 0^{++}$ hypothesis is favored over $2^{++}$ with 2.5$\sigma$. The properties of the new state $X^{*}(3860)$  are well matched to expectations for the $\chi_{c0}(2P)$ resonance. So it seems to be a better candidate for the $\chi_{c0}(2P)$ charmonium state than the $X(3915)$.

\section{Charm Spectroscopy}

\begin{figure}[t]
\centering
\includegraphics[width=67.5mm]{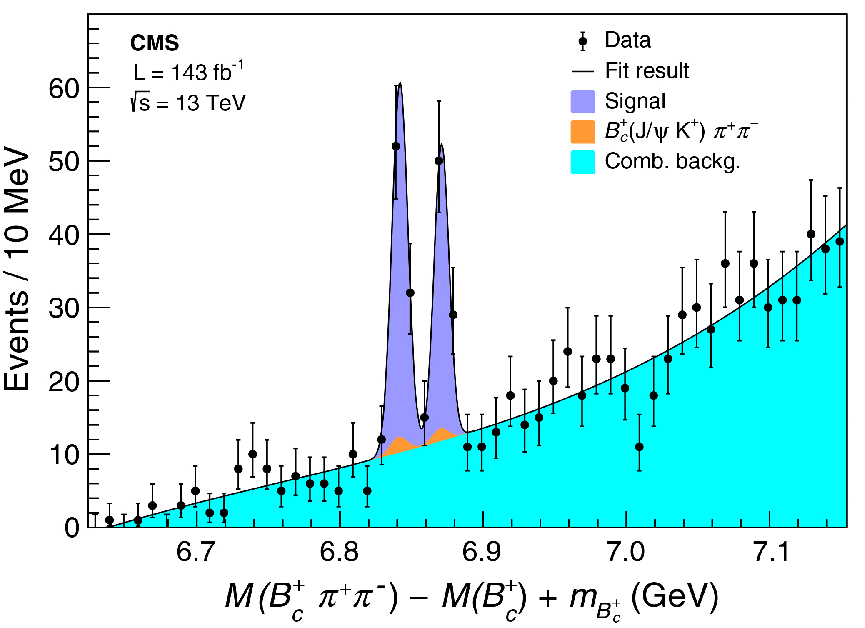}
\includegraphics[width=71.5mm]{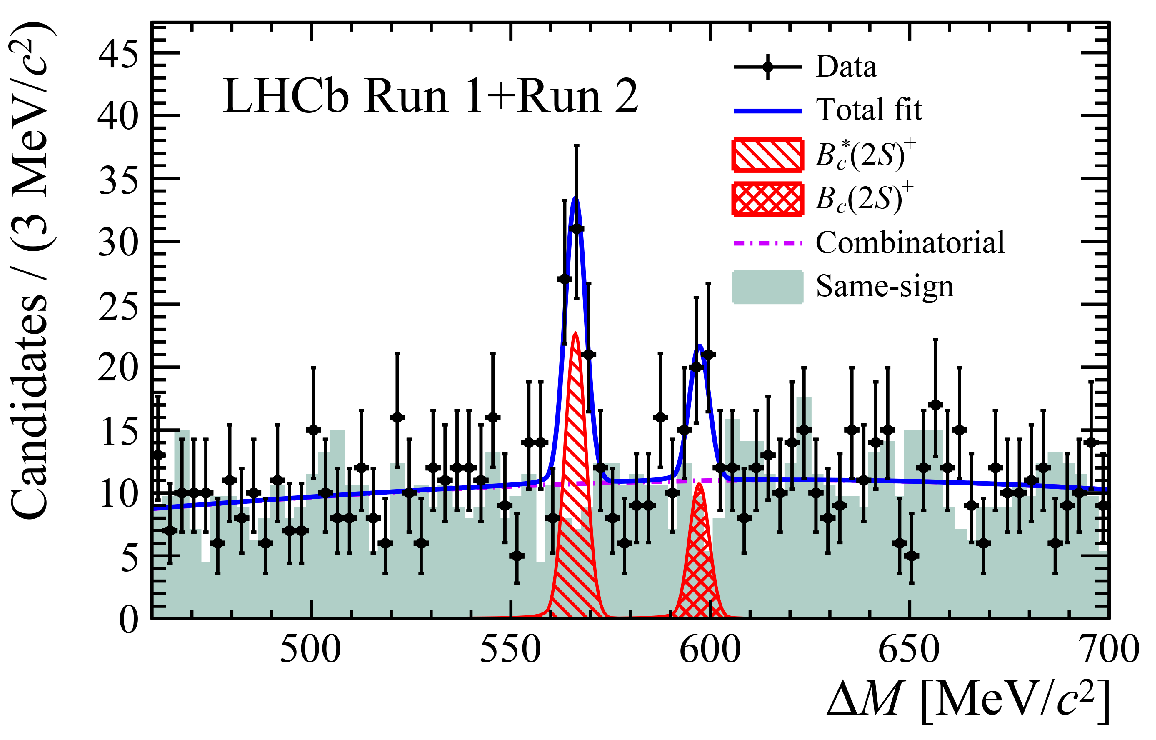}
\caption{ The $M(B_{c}^{+}\pi^{+}\pi^{-}) - M(B_{c}^{+}) + m_{B^{+}_{c}}$ distribution with the fit results overlaid from CMS result (top). The Distribution of $\Delta M \equiv M(B^{+}_{c}\pi^{+}\pi^{-}) - M(B_{c}^{+})$ with the fit results overlaid from LHCb result (bottom).} \label{B_c_CMS_LHCb}
\end{figure}

\subsection{$B_{c}$ spectroscopy}

Since the $\bar{b}c$ mesons cannot annihilate into gluons, the excited states  below $BD$ threshold can only undergo radiative or pionictransitions to the ground state $B_{c}^{+}$. In 2014, ATLAS reported a new resonance in the $B_{c}^{+}(J/\psi\pi^{+})\pi^{+}\pi^{-}$ mass spectrum with mass 6842 $\pm$ 4 $\pm$ 5 MeV/$c^{2}$~\cite{Bc_Atlas}. But due to large mass resolution and low signal yield, no determination could be made as to whether the observed peak is either the $B_{c}(2S)^{+}$, the $B^{*}_{c}(2S)^{+}$ state, or a combination of the two states.

\begin{figure}[t]
\centering
\begin{overpic} [width=39.3mm]{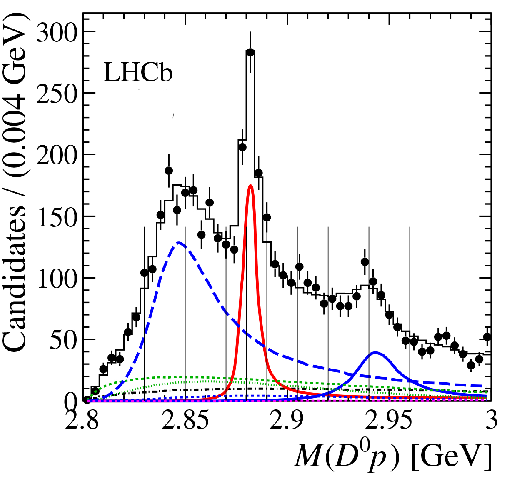}
\put(80,80){(a)}
\end{overpic}
\begin{overpic}[width=40.3mm]{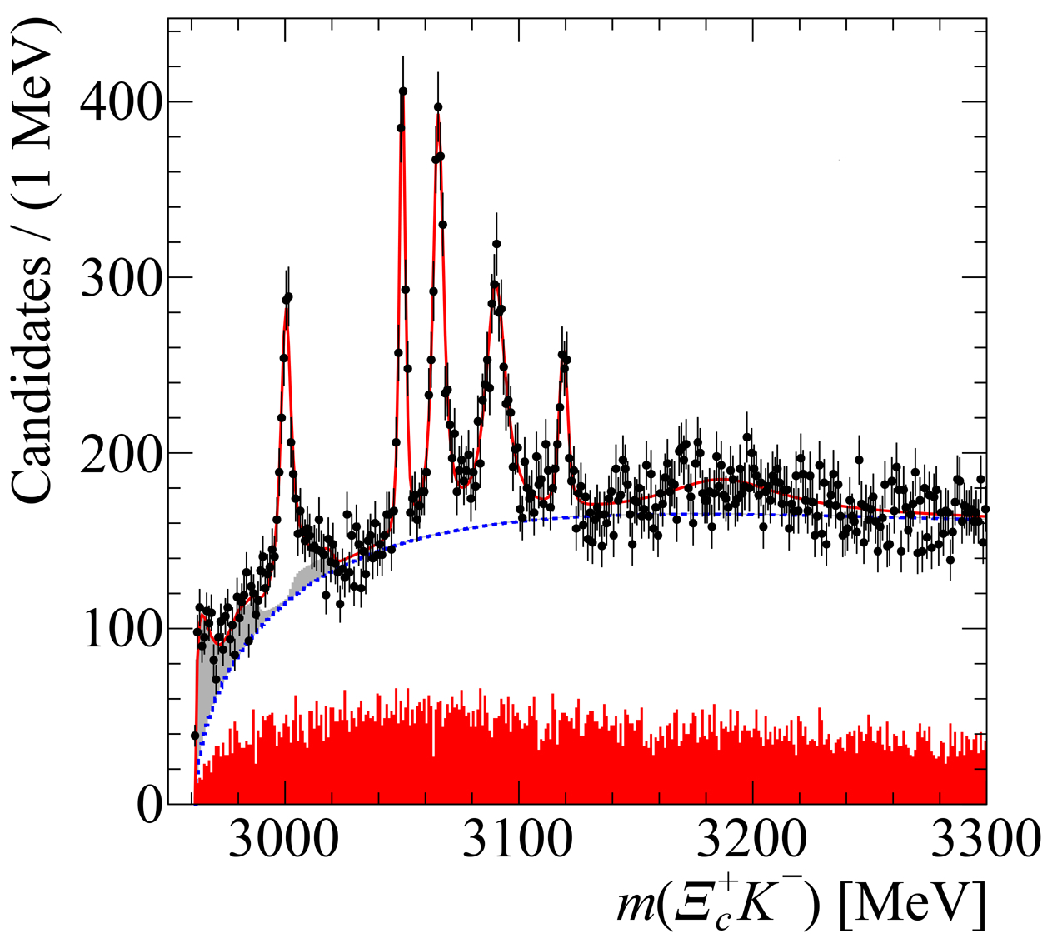}
\put(80,78){(b)}
\end{overpic}
\begin{overpic}[width=39.5mm]{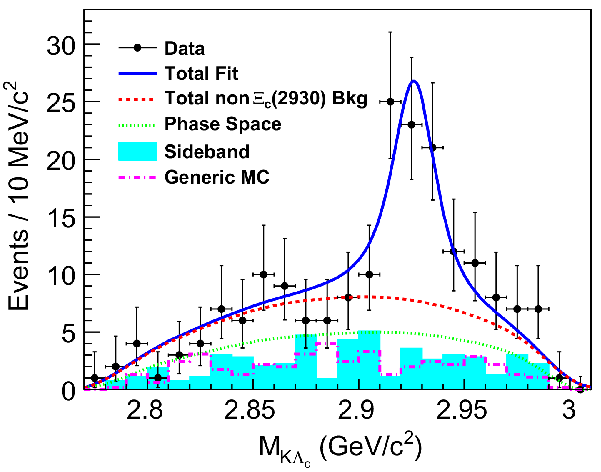}
\put(80,68){(c)}
\end{overpic}
\begin{overpic}[width=40.5mm]{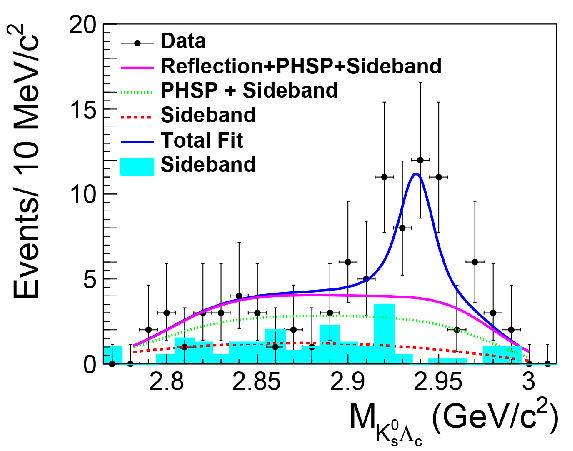}
\put(80,66){(d)}
\end{overpic}
\caption{Results of the fit of the $\Lambda_{b}^{0} \to D^{0}p\pi^{-}$ data in the $D^{0}p$ mass region including the $\Lambda_{c}(2880)^{+}$ and $\Lambda_{c}(2940)^{+}$ resonances (a). Distribution of the reconstructed invariant mass $m(\Xi^{+}_{c} K^{-})$ for all candidates passing
the likelihood ratio selection (b).  The invariant mass distribution of $K^{-}\Lambda_{c}^{+}$ (c) and $K^{0}_{S}\Lambda_{c}^{+}$ (d).} \label{charm_baryons}
\end{figure}

Recently, the CMS collaboration reported the observation of the $B_{c}(2S)^{+}$ and $B^{*}_{c}(2S)^{+}$ states in the $B_{c}^{+}(J/\psi\pi^{+})\pi^{+}\pi^{-}$ mass spectrum with the event sample corresponding to an integrated luminosity of 143 fb$^{-}$ recorded at $\sqrt{s}$ = 13 TeV~\cite{Bc_CMS}. $B_{c}(2S)^{+}$ and $B^{*}_{c}(2S)^{+}$ states are resolved for the first time, sparated in mass by 29.1 $\pm$ 1.5 $\pm$ 0.7 MeV/$c^{2}$ as shown in Fig.~\ref{B_c_CMS_LHCb}.  The $B_{c}(2S)^{+}$ mass is measured to be 6871 $\pm$ 1.2(stat) $\pm$ 0.8(syst) $\pm$ 0.8($B_{c}^{+}$) MeV/$c^{2}$, where the last term is due to the uncertainty in the world-average $B_{c}^{+}$ mass. Because of the low-energy photon emitted in the intermediate radiative decay $B_{c}^{*+} \to B_{c}^{+}\gamma$ is not reconstructed, the observed $B_{c}^{*}(2S)^{+}$ peak has a mass lower than the true value, which remains unknown. The mass difference $M(B_{c}^{*}(2S)^{+})-M(B_{c}^{*+})$ is measured as 567.0 $\pm$ 1.0 (total) MeV.

With the data sample corresponding to an integrated luminosity of 8.5 fb$^{-1}$ recorded at $\sqrt{s}$ = 7, 8 and 13 TeV, LHCb also investigated the excited $B_{c}^{+}$ state in the $B^{+}_{c}\pi^{+}\pi^{-}$ invariant-mass spectrum~\cite{Bc_LHCb}. A  peaking structure consistent with the $B_{c}^{*}(2S)^{+}$ state is observed in the $B^{+}_{c}\pi^{+}\pi^{-}$ mass spectrum as shown in Fig.~\ref{B_c_CMS_LHCb}. The associated mass $M(B_{c}^{*}(2S)^{+})_{rec} = M(B_{c}^{*}(2S)^{+}) - (M(B^{*+}_{c}) - M(B_{c}^{+}))$ is measured as 6841.2 $\pm$ 0.6(stat) $\pm$ 0.1(syst) $\pm$ 0.8($B_{c}^{+}$) MeV/$c^{2}$. A hint for a second structure consistent with the $B_{c}(2S)^{+}$ state is also observed. Its mass is measured to be 6872.1 $\pm$ 1.3(stat) $\pm$ 0.1(syst) $\pm$ 0.8($B_{c}^{+}$) MeV/$c^{2}$. The mass difference of the two $B^{(*)}_{c}(2S)^{+}$ peaks is determined to be 31.0 $\pm$ 1.4 $\pm$ 0.0  MeV/$c^{2}$.

\subsection{Charmed baryons}

\begin{figure}[t]
\centering
\includegraphics[width=64.mm]{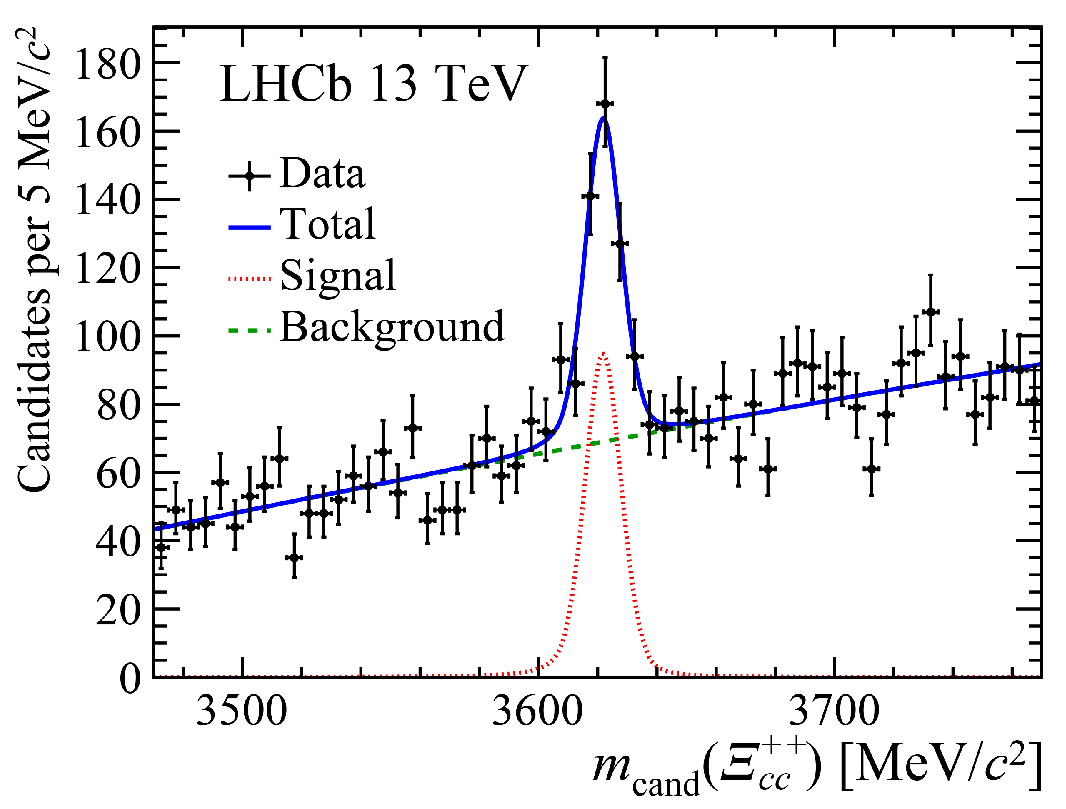}
\includegraphics[width=64.1mm]{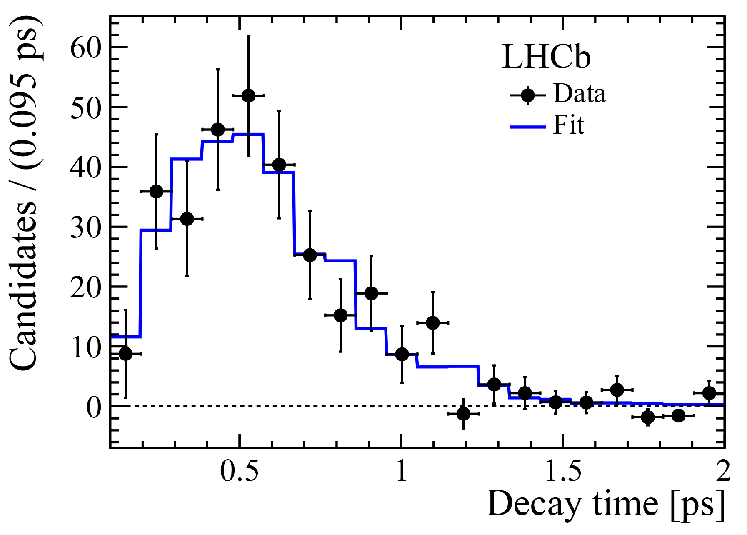}
\caption{Invariant mass distribution of $\Lambda^{+}_{c}K^{-}\pi^{+}\pi^{+}$ candidates with fit projections overlaid (top). Background-subtracted decay-time distribution of selected $\Xi^{++}_{cc} \to \Lambda^{+} K^{-}\pi^{+}\pi^{+}$ candidates (bottom).} \label{Xi_cc_LHCb}
\end{figure}

\begin{table*}[t]
\begin{center}
\caption{Results of the fit to $m(\Xi^{+}_{c}K^{-})$ for the mass, width from LHCb data and the mass from Belle data.}
\begin{tabular}{|l|c|c|c|}
\hline \textbf{Resonance} & \textbf{Mass (MeV)} & \textbf{F (MeV)} &
\textbf{Mass$_{\rm Belle}$ (MeV)}
\\
\hline $\Omega_{c}(3000)^{0}$ & 3000.4 $\pm$ 0.2 $\pm$ 0.1$^{+0.3}_{-0.5}$ & 4.5 $\pm$ 0.6 $\pm$ 0.3  & 3000.7  $\pm$ 1.0 $\pm$ 0.2 \\
\hline  $\Omega_{c}(3050)^{0}$ & 3050.2 $\pm$ 0.1 $\pm$ 0.1$^{+0.3}_{-0.5}$ & 0.8 $\pm$ 0.2 $\pm$ 0.1 & 3050.2 $\pm$ 0.4 $\pm$ 0.2 \\
\hline  $\Omega_{c}(3066)^{0}$ & 3065.6 $\pm$ 0.1 $\pm$ 0.3$^{+0.3}_{-0.5}$ & 3.5 $\pm$ 0.4 $\pm$ 0.2 & 3064.9 $\pm$ 0.6 $\pm$ 0.2 \\
\hline  $\Omega_{c}(3090)^{0}$ & 3090.2 $\pm$ 0.3 $\pm$ 0.5$^{+0.3}_{-0.5}$ & 8.7 $\pm$ 1.0 $\pm$ 0.8 & 3089.3 $\pm$ 1.2 $\pm$ 0.2 \\
\hline  $\Omega_{c}(3119)^{0}$ & 3119.1 $\pm$ 0.3 $\pm$ 0.9$^{+0.3}_{-0.5}$ & 1.1 $\pm$ 0.8 $\pm$ 0.4 & - \\
\hline  $\Omega_{c}(3188)^{0}$ & 3188 $\pm$ 5 $\pm$ 13 & 60 $\pm$ 15 $\pm$ 11  & 3199 $\pm$ 9 $\pm$ 4 \\
\hline
\end{tabular}
\label{Omegac_mass}
\end{center}
\end{table*}

Recently, LHCb performed an amplitude analysis of the decay $\Lambda_{b}^{0} \to D^{0} p \pi^{-}$ with the data sample corresponding to an integrated luminosity of 3.0 fb$^{-1}$ recorded at $\sqrt{s}$ = 7 and 8 TeV~\cite{Dp_LHCb}. The spectroscopy of excited $\Lambda^{+}$ states in the $D^{0}p$ amplitude are studied in detail as shown in Fig.~\ref{charm_baryons}. The preferred spin of the $\Lambda_{c}(2880)^{+}$ state is found to be $J=5/2$. The mass and width of the $\Lambda_{c}(2880)^{+}$ are measured to be 2881.75 $\pm$ 0.29(stat) $\pm$ 0.07(syst)$^{+0.14}_{-0.20}$(model) MeV/$c^{2}$ and 5.43$^{+0.77}_{-0.71}$(stat) $\pm$ 0.29(syst)$^{+0.75}_{-0.00}$(model) MeV, respectively. The last uncertainty is due to the modelling of the nonresonant amplitudes. The spin and parity of the $\Lambda_{c}(2940)^{+}$ state are firstly determined as $J^{P} = 3/2^{-}$, which is consistent with its interpretations as a $D^{*}N$ molecule~\cite{D*N_2940} or a radial 2$P$ excitation~\cite{radial2P_2940}, but the other solutions with spins 1/2 to 7/2 cannot be excluded. An enhancement near the threshold of $D^{0}p$ amplitude is observed with the mass and width of  2856.1$^{+2.0}_{-1.7}$(stat) $\pm$ 0.5(syst)$^{+1.1}_{-5.6}$(model) MeV/$c^{2}$ and 67.6$^{+10.1}_{-8.1}$(stat) $\pm$ 1.4(syst)$^{+5.9}_{-20.0}$(model) MeV. The quantum numbers are determined as $J^{P} = 3/2^{+}$, with the parity measured relative to that of the $\Lambda_{c}(2880)^{+}$ state. The mass of the $\Lambda_{c}(2860)^{+}$ state is consistent with predictions for an orbital D-wave excited $\Lambda^{+}_{c}$  with quantum numbers $3/2^{+}$ based on the nonrelativistic heavy quark-light diquark model~\cite{diquark_2860} and from QCD sum rules in the HQET framework~\cite{HQET_2860}.

The $\Xi_{c}(2930)$ charmed-strange baryon has only been reported by BaBar in the analysis of $B^{-} \to K^{-}\Lambda^{+}_{c}\bar{\Lambda}^{-}$~\cite{Xi0_Klambda_BaBar}. A signal with a mass of 2931 $\pm$ 3 $\pm$ 5 MeV/$c^{2}$ and a width of 36 $\pm$ 7 $\pm$ 11 MeV in $K^{-}\Lambda^{+}_{c}$  mass spectrum was claimed. However, neither the fit result to the  spectrum nor the significance of the signal were given. Recently, Belle  performed  an updated measurement of $B^{-} \to K^{-}\Lambda^{+}_{c}\bar{\Lambda}^{-}$ with its full $\Upsilon(4S)$ data sample corresponding to an integrated luminosity of 711 fb$^{-1}$~\cite{Xi0_Klambda_Belle}. The $\Xi_{c}(2930)^{0}$ signal is observed with a significance of 5.1$\sigma$ as shown in Fig.~\ref{charm_baryons}. Its mass and width are measured as 2928.9 $\pm$ 3.0$^{+0.9}_{-12.0}$ MeV/$c^{2}$ and 19.5 $\pm$ 8.4$^{+5.9}_{-7.9}$ MeV, respectively. Belle also investigate the $\Xi_{c}(2930)^{+}$ in the process $B^{0} \to K^{0}_{S}\Lambda^{+}_{c}\bar{\Lambda}^{-}$~\cite{Xip_KSlambda_Belle}. Only an evidence of the $\Xi_{c}(2930)^{+}$ with a 4.1$\sigma$ is observed in $K^{0}_{S}\Lambda^{+}_{c}$ mass spectrum as shown in Fig.~\ref{charm_baryons}. Its mass and width are measured as 2942.3 $\pm$ 4.4 $\pm$ 1.5 MeV/$c^{2}$ and 14.8 $\pm$ 8.8 $\pm$ 2.5 MeV, respectively.

Excited $\Lambda_{c}$, $\Sigma_{c}$, $\Xi_{c}$ states both have been reported, but no excited $\Omega_{c}$ states were observed before LHCb. Recently,  LHCb searched for new $\Omega_{c}^{*0}$ states that decay strongly to the final state $\Xi_{c}^{+}K^{-}$, base on the data sample corresponding to an integrated luminosity of 3.3 fb$^{-1}$ recorded at $\sqrt{s}$ = 7, 8 and 13 TeV~\cite{Omega_LHCb}. Five new, narrow excited $\Omega_{c}^{0}$ states are observed as  shown in Fig.~\ref{charm_baryons}: the $\Omega_{c}(3000)^{0}$, $\Omega_{c}(3050)^{0}$, $\Omega_{c}(3066)^{0}$, $\Omega_{c}(3090)^{0}$, and $\Omega_{c}(3119)^{0}$. Further more, the data indicate also the presence of a broad structure around 3188 MeV. The masses and widthes of those new states are measured as shown in Table~\ref{Omegac_mass}. Soon later, Belle also measured the same process with its full data sample corresponding to an integrated luminosity of 980 fb$^{-1}$~\cite{Omega_Belle}. All the $\Omega_{c}^{*0}$ except $\Omega_{c}(3119)^{0}$ are confirmed. With the width of $\Omega_{c}^{*0}$ states fixed to the values from LHCb result, the masses of those states are measured as list in the Table~\ref{Omegac_mass}.

The existence of doubly charmed baryons predicted by quark model~\cite{quark_model}. Three weakly decaying $qqq$ states with C = 2 are expected: one isospin doublet ($\Xi^{++}_{cc} = ccu$ and $\Xi^{+}_{cc} = ccd$) and one isospin singlet ($\Omega^{+}_{cc} = ccs$), each with spin-parity $J^{P} = 1/2^{+}$. The masses of the $\Xi_{cc}$ states are predicted to lie in the range 3500 to 3700 MeV/$c^{2}$~\cite{Xi_cc_mass1,Xi_cc_mass2}.  The observation of $\Xi^{++}_{cc}$ has been claimed by SELEX~\cite{Xi_cc_SELEX1,Xi_cc_SELEX2}.  But no evidence observed by BaBar~\cite{Xi_cc_BaBar}, FOCUS~\cite{Xi_cc_FOCUS}, Belle~\cite{Xi_cc_Belle} and LHCb~\cite{Xi_cc_LHCb_old}. Recently, LHCb searched for $\Xi^{++}_{cc}$ via the most promising channel $\Xi^{++}_{cc}\to\Lambda_{c}^{+}K^{-}\pi^{+}\pi^{-}$ base on the data sample corresponding to an integrated luminosity of 1.7 fb$^{-1}$ recorded at $\sqrt{s}$ = 13 TeV~\cite{Xi_cc_LHCb}. $\Xi^{++}_{cc}$ is observed for the first time as shown in Fig~\ref{Xi_cc_LHCb}, and confirmed in an additional sample of data collected at 8 TeV.  The mass of the structure is measured to be 3621 $\pm$ 0.72(stat) $\pm$ 0.27(syst) $\pm$ 0.14($\Lambda_{c}^{+}$) MeV/$c^{2}$. The last uncertainty is due to the limited knowledge of the $\Lambda_{c}^{+}$ mass. Soon later, using the same date with extra trigger requirement, the decay-time distribution of $\Xi^{++}_{cc}$ are measured relative to $\Lambda^{0}_{b}\to\Lambda_{c}^{+}K^{-}\pi^{+}\pi^{-}$ as shown in Fig.~\ref{Xi_cc_LHCb}~\cite{Xi_cc_lifetime_LHCb}. The $\Xi^{++}_{cc}$ lifetime is measured to be 0.256$^{+0.024}_{-0.022}$(stat) $\pm$ 0.014(syst) ps, which establishes the weakly decaying nature of the recently discovered  $\Xi^{++}_{cc}$.

\section{Summary}

There is a wide range of interesting charmonium and charm spectroscopy results, but only a small selection of recent results has been presented in this talk.
The Measurements of resonance parameters of $\chi_{c1}$, $\chi_{c2}$, $\eta_{c}$, $\eta_{c}(2S)$, and $\psi(3770)$  were improved.
 Candidate for $\psi(1^{3}D_{3})$ and alternative candidate for $\chi_{c0}(2P)$ were  observed.  Excited $B_{c}$ states, candidate for D-wave excited $\Lambda_{c}$ states, excited $\Xi_{c}$, $\Omega_{c}$ states, and doubly charmed baryon $\Xi_{cc}^{++}$ were observed. The lifetime of $\Xi_{cc}^{++}$ was measured, which establishes the weakly decaying nature of $\Xi_{cc}^{++}$. BESIII will keep taking data in the region of charmonium. Belle II just started its Phase III data taking. With the upgrade, LHCb will get much more data.  Many more precise spectroscopic measurements are expected in the near feature.

\bigskip 

\begin{thebibliography}{9}   

\bibitem{JPsi1974_1} J. J. Aubert {\em et al.},  Phys. Rev. Lett. {\bf 33} 1404 (1974).
\bibitem{JPsi1974_2} J. E. Augustin {\em et al.},  Phys. Rev. Lett. {\bf 33} 1406 (1974).
\bibitem{psip1974} G. S. Abrams {\em et al.}, Phys. Rev. Lett. {\bf 33} 1453 (1974).
\bibitem{psipp1977} P. A. Rapidis {\em et al.}, Phys. Rev. Lett. {\bf 39} 974 {1977}.
\bibitem{X3872_Belle} S.-K. Choi {\em et al.} (Belle Collaboration), Phys. Rev. Lett. {\bf 91} 262001 (2003).
\bibitem{XYZreview1} N. Brambilla  {\em et al.}, Eur. Phys. J. C  {\bf 74}, 2981 (2014).
\bibitem{XYZreview2} A. J. Bevan {\em et al.} (BaBar and Belle Collaborations), Eur. Phys. J. C 74, 3026 (2014).
\bibitem{XYZreview3}  C. Z. Yuan, Int. J. Mod. Phys. A {\bf 33}, no. 21, 1830018 (2018).


\bibitem{Bc_CDF} F. Abe {\em et al.} (CDF Collaboration), Phys. Rev. Lett. {\bf 81}  2432 (1998).
\bibitem{Bc_pre_1} S. Narison, Phys. Lett. B {\bf 210}, 238 (1988);
\bibitem{Bc_pre_2} E. J. Eichten and C. Quigg, Phys. Rev. D {\bf 49}, 5845 (1994);
\bibitem{Bc_pre_3} D. Ebert, R. N. Faustov, and V. O. Galkin, Phys. Rev. D {\bf 67}, 014027 (2003);
\bibitem{Bc_pre_4} R. J. Dowdall, C. T. H. Davies, T. C. Hammant, and R. R. Horgan, Phys. Rev. D {\bf 86}, 094510 (2012).

\bibitem{Chcj_Jpsimumu_LHCb} R. Aaij {\em et al.} (LHCb Collaboration), Phys. Rev. Lett. {\bf 119}, 221801 (2017).
\bibitem{PDG2018} M. Tanabashi {\em et al.} (Particle Data Group), Phys. Rev. D {\bf 98}, 030001 (2018).
\bibitem{Chcj_Jpsimumu_BESIII} M. Ablikim {\em et al.} (BESIII Collaboration), Phys. Rev. D {\bf 99}, 051101(R) (2019).
\bibitem{Etac_PP_LHCb} R. Aaij {\em et al.} (LHCb Collaboration), Phys. Lett. B {\bf 769}, 305 (2017).
\bibitem{Etac_phiphi_LHCb} R. Aaij {\em et al.} (LHCb Collaboration), Eur. Phys. J. C  {\bf 77}: 609 (2017).
\bibitem{Etac_PP_LHCb_ex}  R. Aaij {\em et al.} (LHCb collaboration), Phys. Lett. B {\bf 769} 305 (2017).
\bibitem{psi3D3_LHCb} R. Aaij {\em et al.} (LHCb Collaboration), arXiv:1903.1224 0.
\bibitem{TBarnes} T. Barnes, S. Godfrey, and E. S. Swanson Phys. Rev. D {\bf 72}, 054026 (2005).

\bibitem{X3915_B_jpsiOmega_BaBar} B. Aubert {\em et al.} (BABAR Collaboration), Phys. Rev. Lett. {\bf 101}, 082001 (2008).
\bibitem{X3915_B_jpsiOmega_Belle} S.-K. Choi {\em et al.} (Belle Collaboration), Phys. Rev. Lett. {\bf 94}, 182002 (2005).
\bibitem{X3915_gammagamma_jpsiOmega_Belle} S. Uehara {\em et al.} (Belle Collaboration), Phys. Rev. Lett. {\bf 104}, 092001 (2010).
\bibitem{X3915_gammagamma_jpsiOmega_BaBar} J. P. Lees {\em et al.} (BABAR Collaboration), Phys. Rev. D {\bf 86}, 072002 (2012).
\bibitem{Chic02P_width} E. J. Eichten, K. Lane and C. Quigg, Phys. Rev. D {\bf 69}, 094019 (2004).
\bibitem{Barnes} Barnes, S. Godfrey, and E. S. Swanson, Phys. Rev. D {\bf 72}, 644 054026 (2005).


\bibitem{Chic02P_Belle} K. Chilikin {\em et al.} (Belle Collaboration), Phys. Rev. D {\bf 95}, 112003 (2017).
\bibitem{Chic02P_Theory} D. Ebert, R. N. Faustov and V. O. Galkin, Phys. Rev. D {\bf 67}, 014027 (2003).



\bibitem{Bc_Atlas} G. Aad {\em et al.} (ATLAS Collaboration), Phys. Rev. Lett. {\bf 113}, 212004 (2014).
\bibitem{Bc_CMS} A. M. Sirunyan {\em et al.} (CMS Collaboration), Phys. Rev. Lett. {\bf 122}, 132001 (2019).
\bibitem{Bc_LHCb} R. Aaij {\em et al.} (LHCb Collaboration), arXiv:1904.000 81.

\bibitem{Dp_LHCb} Aaij, R., Adeva, B. {\em et al.}, J. High Energ. Phys. (2017) {\bf 2017}: 30.
\bibitem{D*N_2940} J.-R. Zhang,  Phys. Rev. D {\bf 89} 096006 (2014).
\bibitem{radial2P_2940} B. Chen, K.-W. Wei, and A. Zhang, Eur. Phys. J. A {\bf 51} 82 (2015).
\bibitem{diquark_2860} B. Chen, K.-W. Wei, X. Liu, and T. Matsuki, arXiv:1609.07967.
\bibitem{HQET_2860} H.-X. Chen {\em et al.},  Phys. Rev. D {\bf 94} 114016 (2016).

\bibitem{Xi0_Klambda_BaBar} B. Aubert {\em et al.} (BaBar Collaboration), Phys. Rev. D {\bf 77}, 031101 (2008).
\bibitem{Xi0_Klambda_Belle}  Y. B. Li {\em et al.} (Belle Collaboration), Eur. Phys. J. C (2018) {\bf 78}: 252.
\bibitem{Xip_KSlambda_Belle}  Y. B. Li {\em et al.} (Belle Collaboration), Eur. Phys. J. C (2018) {\bf 78}: 928.

\bibitem{Omega_LHCb} R. Aaij {\em et al.} (LHCb Collaboration), Phys. Rev. Lett. {\bf 118}, 182001 (2017).
\bibitem{Omega_Belle} J. Yelton {\em et al.} (Belle Collaboration), Phys. Rev. D {\bf 97}, 051102(R) (2018).

\bibitem{quark_model} M. Gell-Mann,  Phys. Lett. {\bf 8} 214 (1964).
\bibitem{Xi_cc_mass1} D.-H. He et al.,  Phys. Rev. D {\bf 70} 094004 (2004).
\bibitem{Xi_cc_mass2} C.-H. Chang, C.-F. Qiao, J.-X. Wang, and X.-G. Wu, Phys. Rev. D {\bf 73}  094022 (2006).
\bibitem{Xi_cc_SELEX1}  M. Mattson {\em et al.} (SELEX collaboration), Phys. Rev. Lett. {\bf 89}  112001 (2002).
\bibitem{Xi_cc_SELEX2}  A. Ocherashvili {\em et al.} (SELEX collaboration), Phys. Lett. B {\bf 628}  18 (2005).
\bibitem{Xi_cc_BaBar} B. Aubert {\em et al.} (BaBar collaboration), Phys. Rev. D {\bf 74}  011103 (2006).
\bibitem{Xi_cc_FOCUS} S. P. Ratti,  Nucl. Phys. Proc. Suppl. {\bf 115} 33 (2003).

\bibitem{Xi_cc_Belle} R. Chistov {\em et al.} (Belle collaboration),  Phys. Rev. Lett. {\bf 97}  162001 (2006).
\bibitem{Xi_cc_LHCb_old} R. Aaij {\em et al.} (LHCb collaboration), JHEP {\bf 12}  090 (2013).
\bibitem{Xi_cc_LHCb} R. Aaij {\em et al.} (LHCb Collaboration), Phys. Rev. Lett. {\bf 119}, 112001 (2017).
\bibitem{Xi_cc_lifetime_LHCb} R. Aaij {\em et al.} (LHCb Collaboration), Phys. Rev. Lett. {\bf 121}, 052002 (2018).







\end{thebibliography}

\end{document}